\documentstyle[12pt]{article}
\begin{document}
\title{The Magnetization by the
Finite Energy Constant Magnetic Field in the Two -- Dimensional Ising Model.}

\author{Yury M. Zinoviev\thanks{This paper is the corrected version of the paper: Yury M. Zinoviev, The
Finite Energy Constant Magnetization of the Two -- Dimensional Ising Model. cond --
mat/0108489
 This work is supported in part by the Russian Foundation for Basic Research
(Grant No. 00 -- 01 -- 00083)}\\ Steklov Mathematical Institute,\\
Gubkin St. 8, Moscow 117966, GSP -- 1, Russia, \\ e -- mail: zinoviev@mi.ras.ru}

\date{}

\maketitle

\vskip 1cm

\noindent {\bf Abstract.} We suggest the new definition of the magnetization. For the
two -- dimensional Ising model with the free boundary conditions we calculate this
magnetization.

\vskip 1cm

\section{Introduction}

\noindent Yang \cite{1} proved the formula for the spontaneous magnetization
of the two -- dimensional Ising model. (See also the paper \cite{2} where
the result for different vertical and horizontal interactions is obtained.)
Montroll, Potts and Ward \cite{3} defined the square of the spontaneous
magnetization as the asymptotics of the spin -- spin correlation function.
In order to characterize these papers we cite the paper \cite{4}:

\noindent "In contrast to the free energy, the spontaneous magnetization of
the Ising model on a square lattice, correctly defined, has never been solved
with complete mathematical rigor. Starting from the only sensible definition
of the spontaneous magnetization, the methods of Yang, and of Montroll,
Potts, and Ward are each forced to make an assumption that has not been
rigorously justified."

In the book (\cite{5}, p. 113) the following problem is solved:

\noindent "More precisely, we impose on the Onsager lattice a cyclic
boundary condition in the horizontal direction only and let a magnetic field
interact with one of the two horizontal boundary rows of spins."

In the paper \cite{6} the estimates of the derivatives of the magnetization with
respect to the magnetic field for zero magnetic field are obtained. These estimates are
obtained for the domain where the interaction energy divided by the temperature is
sufficiently large. We will study the magnetization for the domain where the
interaction energy divided by the temperature is sufficiently small.

In this paper we suggest the new definition of the magnetization. In order
to formulate it we use the one -- dimensional Ising model. Let for any
integer $k = 1,...,N$ the number $\sigma_{k} = \pm 1$ be given. The energy
for the Ising model with the magnetic field is
\begin{equation}
\label{1.1}
\bar{H} (\sigma) = - \sum_{k\, =\, 1}^{N\, -\, 1} E\sigma_{k} \sigma_{k + 1}
- E\sigma_{N} \sigma_{1} - \sum_{k\, =\, 1}^{N} H(k)\sigma_{k}
\end{equation}
where the number $H(k) = H$ for any integer $k = 1,...,N$. The function
\begin{equation}
\label{1.2}
Z_{N}(E,H) = \sum_{{\sigma_{k} \, =\, \pm \, 1,}\atop {k\, =\, 1,...,N}}
\exp \{ - \beta \bar{H} (\sigma )\}
\end{equation}
is called the partition function of the Ising model. Due to (\cite{5},
Chapter III, formulae (2.10), (2.13))
\begin{eqnarray}
\label{1.3}
Z_{N}(E,H) = (\lambda_{+} (E,H))^{N} + (\lambda_{-} (E,H))^{N}, \nonumber \\
\lambda_{\pm} (E,H) = e^{\beta E}
(\cosh \beta H \pm (\sinh^{2} \beta H + e^{- 4\beta E})^{1/2}).
\end{eqnarray}
Due to (\cite{5}, Chapter II, formula (4.5)) the average total magnetization
is
\begin{equation}
\label{1.4}
\bar{M}_{N} (E,H) = \beta^{- 1} \frac{\partial}{\partial H} \ln Z_{N}(E,H).
\end{equation}
The relations (\ref{1.3}) imply
\begin{equation}
\label{1.5}
\bar{M}_{N} (E,H) = NM(E,H) + o(N)
\end{equation}
where the coefficient
\begin{equation}
\label{1.6}
M(E,H) = (\sinh^{2} \beta H + e^{- 4\beta E})^{- 1/2}\sinh \beta H
\end{equation}
is called the magnetization per spin. Since $M(E,0) = 0$, the spontaneous
magnetization is absent in this model.

The energy of the magnetic field in the expression (\ref{1.1}) is
\begin{equation}
\label{1.7}
\sum_{k\, =\, 1}^{N} (H(k))^{2} = NH^{2}.
\end{equation}
This energy tends to the infinity when $N \rightarrow \infty $. Hence the
above problem seems to be non -- physical one. We change the numbers
$H(k) = H$ in the expression (\ref{1.1}) for the numbers $H(k) = N^{- 1/2}H$.
The energy (\ref{1.7}) of this magnetic field is constant when
$N \rightarrow \infty $. The relations (\ref{1.3}) imply
\begin{equation}
\label{1.8}
m(E,H) = \lim_{N \rightarrow \infty }
\beta^{- 1} \frac{\partial}{\partial H} \ln Z_{N}(E,N^{- 1/2}H) =
\beta He^{2\beta E}.
\end{equation}
The quantity (\ref{1.8}) is called the magnetization by the finite energy constant
magnetic field in the one -- dimensional Ising model.

In this paper we aim to calculate the magnetization by the finite energy constant
magnetic field in the two -- dimensional Ising model with the free boundary conditions.

\section{Magnetization}
\setcounter{equation}{0}

\noindent We consider a rectangular lattice on the plane formed by the points
with the integral Cartesian coordinates $x = k_{1}$, $y = k_{2}$,
$M^{\prime}_{i} \leq k_{i} \leq M_{i}$, $i = 1,2$, and the corresponding
horizontal and vertical edges connecting the neighbour vertices.
We denote this graph by $G(M^{\prime}_{1},M^{\prime}_{2};M_{1},M_{2})$.
The cell complex $P(G)$ is called the set consisting of the cells
(vertices, edges, faces). A vertex of $P(G)$ is called a cell
of dimension $0$. It is denoted by $s^{0}_{i}$. An edge of $P(G)$ is
called a cell of dimension $1$. It is denoted by $s^{1}_{i}$. A face of
$P(G)$ is called a cell of dimension $2$. It is denoted by $s^{2}_{i}$.
We denote by ${\bf Z}_{2}^{add}$ the group of modulo $2$ residuals. The
modulo $2$ residuals are multiplied by each other and the group
${\bf Z}_{2}^{add}$ is a field. To every pair of the cells
$s^{p}_{i}, s^{p - 1}_{j}$ there corresponds the number
$(s^{p}_{i}:s^{p - 1}_{j}) \in {\bf Z}^{add}_{2}$ (incidence number).
If the cell $s^{p - 1}_{j}$ is included into the boundary of the cell
$s^{p}_{i}$, then the incidence number $(s^{p}_{i}:s^{p - 1}_{j}) = 1$.
Otherwise the incidence number $(s^{p}_{i}:s^{p - 1}_{j}) = 0$. For any
pair of the cells $s^{2}_{i},s^{0}_{j}$ the incidence numbers satisfy
the condition
\begin{equation}
\label{2.1}
\sum_{m} (s^{2}_{i}:s^{1}_{m})(s^{1}_{m}:s^{0}_{j}) = 0 \,
\hbox{mod} \, 2.
\end{equation}
Indeed, if the vertex $s^{0}_{j}$ is not included into the boundary of
the square $s^{2}_{i}$ , then the condition (\ref{2.1}) is fulfilled.
If the vertex $s^{0}_{j}$ is included into the boundary of the square
$s^{2}_{i}$, then it is included into the boundaries of four edges
$s^{1}_{m}$, two of which are included into the boundary of the square
$s^{2}_{i}$. The condition (\ref{2.1}) is fulfilled again.

A cochain $c^{p}$ of the complex $P(G)$ with the coefficients in the
group ${\bf Z}^{add}_{2}$ is a function on the $p$ -- dimensional cells
taking values in the group ${\bf Z}^{add}_{2}$. Usually the cell
orientation is considered and the cochains are antisymmetric functions:
$c^{p}(- s^{p}_{i}) = - c^{p}(s^{p}_{i})$. However,
$- 1 = 1 \, \hbox{mod} \, 2$ and we can neglect the cell orientation for
the coefficients in the group ${\bf Z}^{add}_{2}$. The cochains form an
Abelian group
\begin{equation}
\label{2.2}
(c^{p} + c^{\prime p})(s^{p}_{i}) = c^{p}(s^{p}_{i}) +
c^{\prime p}(s^{p}_{i}) \, \hbox{mod} \, 2.
\end{equation}
It is denoted by $C^{p}(P(G),{\bf Z}^{add}_{2})$. The mapping
\begin{equation}
\label{2.3}
\partial c^{p}(s^{p - 1}_{i}) = \sum_{j} (s^{p}_{j}:s^{p - 1}_{i})
c^{p}(s^{p}_{j}) \, \hbox{mod} \, 2
\end{equation}
defines the homomorphism of the group $C^{p}(P(G),{\bf Z}^{add}_{2})$
into the group $C^{p - 1}(P(G),{\bf Z}^{add}_{2})$. It is called the
boundary operator. The mapping
\begin{equation}
\label{2.4}
\partial^{\star} c^{p}(s^{p + 1}_{i}) = \sum_{j} (s^{p + 1}_{i}:s^{p}_{j})
c^{p}(s^{p}_{j}) \, \hbox{mod} \, 2
\end{equation}
defines the homomorphism of the group $C^{p}(P(G),{\bf Z}^{add}_{2})$
into the group $C^{p + 1}(P(G),{\bf Z}^{add}_{2})$. It is called the
coboundary operator. The condition (\ref{2.1}) implies
$\partial \partial = 0$, $\partial^{\star} \partial^{\star} = 0$. The
kernel $Z_{p}(P(G),{\bf Z}^{add}_{2})$ of the homomorphism (\ref{2.3})
on the group $C^{p}(P(G),{\bf Z}^{add}_{2})$ is called the group of
cycles of the complex $P(G)$ with the coefficients in the group
${\bf Z}^{add}_{2}$. The image $B_{p}(P(G),{\bf Z}^{add}_{2})$ of
the homomorphism (\ref{2.3}) in the group $C^{p}(P(G),{\bf Z}^{add}_{2})$
is called the group of boundaries of the complex $P(G)$ with the
coefficients in the group ${\bf Z}^{add}_{2}$. Since $\partial \partial = 0$,
the group $B_{p}(P(G),{\bf Z}^{add}_{2})$ is the subgroup of the
group $Z_{p}(P(G),{\bf Z}^{add}_{2})$. Analogously, for the coboundary
operator $\partial^{\star} $ the group of cocyles
$Z^{p}(P(G),{\bf Z}^{add}_{2})$ and the group of coboundaries
$B^{p}(P(G),{\bf Z}^{add}_{2})$ are defined.

It is possible to introduce the bilinear form on
$C^{p}(P(G),{\bf Z}^{add}_{2})$:
\begin{equation}
\label{2.5}
\langle f^{p},g^{p}\rangle = \sum_{i} f^{p}(s^{p}_{i})g^{p}(s^{p}_{i})
\, \hbox{mod} \, 2.
\end{equation}
The definitions (\ref{2.3}), (\ref{2.4}) imply
\begin{eqnarray}
\label{2.6}
\langle f^{p},\partial^{\star} g^{p - 1}\rangle =
\langle \partial f^{p},g^{p - 1}\rangle \nonumber \\
\langle f^{p},\partial g^{p + 1}\rangle =
\langle \partial^{\star} f^{p},g^{p + 1}\rangle .
\end{eqnarray}

Let a cochain $ \sigma \in C^{0}(P(G),{\bf Z}^{add}_{2})$. The function
\begin{equation}
\label{2.7}
\bar{H}(\sigma ) = - \sum_{s^{1}_{i} \in P(G)} E(s_{i}^{1})
(- 1)^{\partial^{\star} \sigma (s^{1}_{i})} -
\sum_{s_{i}^{0} \in P(G)} H(s_{i}^{0}) (- 1)^{\sigma (s_{i}^{0})}
\end{equation}
is called the energy for the Ising model with the magnetic field. The number
$E(s_{i}^{1})$ is the interaction energy attached to the edge $s_{i}^{1}$.
The number $H(s_{i}^{0})$ is the magnetic field attached to the vertex
$s_{i}^{0}$.

The function
\begin{equation}
\label{2.8} Z_{G}(H) = \sum_{\sigma \in C^{0}(P(G),Z^{add}_{2})} \exp \{ - \beta
\bar{H} (\sigma )\}
\end{equation}
is called the partition function of the Ising model with the magnetic field.

Let the cochain $\chi \in C^{0}(P(G),{\bf Z}^{add}_{2})$ take the
value $1$ at the vertices $s^{0}_{1},...,s^{0}_{m}$ and be equal to $0$ on
all other vertices of the graph
$G(M_{1}^{\prime },M_{2}^{\prime };M_{1},M_{2})$. The correlation function
on the vertices $s^{0}_{1},...,s^{0}_{m}$ of the lattice
$G(M_{1}^{\prime },M_{2}^{\prime };M_{1},M_{2})$ is the function
\begin{equation}
\label{2.9} W_{G}(\chi, H) = (Z_{G}(H))^{- 1} \sum_{\sigma \in C^{0}(P(G),Z^{add}_{2})}
(- 1)^{\langle \chi ,\sigma \rangle } \exp \{ - \beta \bar{H} (\sigma )\} .
\end{equation}

Let us consider the particular case of the function (\ref{2.7}) when
the magnetic field $H(s_{i}^{0})$ is independent of the vertex $s_{i}^{0}$.
We suppose that the energy of this magnetic field
\begin{equation}
\label{2.10}
\sum_{s_{i}^{0} \in P(G)} (H(s_{i}^{0}))^{2}
\end{equation}
is independent of the size of the lattice
$G(M_{1}^{\prime },M_{2}^{\prime };M_{1},M_{2})$. Therefore
\begin{equation}
\label{2.11}
H(s_{i}^{0}) = (\# (VG))^{- 1/2}H
\end{equation}
where $\# (VG)$ is the total number of the vertices of the graph
$G(M_{1}^{\prime },M_{2}^{\prime };M_{1},M_{2})$ and $H$ is a constant.

Analogously to the definition (\ref{1.8}) we define the magnetization by the finite
energy constant magnetic field as
\begin{equation}
\label{2.12} m_{G}(H) = \beta^{- 1} \frac{\partial}{\partial H} \ln Z_{G}((\# (VG))^{-
1/2}H).
\end{equation}
It is easy to verify that for $\epsilon = 0,1$
\begin{equation}
\label{2.13}
\exp \{ \beta E(s^{1}_{i})(- 1)^{\epsilon }\} =
(\cosh \beta E(s^{1}_{i}))\sum_{\xi = 0,1}
(- 1)^{\xi \epsilon }
(\tanh \beta E(s^{1}_{i}))^{\frac{1}{2} (1 - (- 1)^{\xi })}.
\end{equation}
The equalities (\ref{2.6}), (\ref{2.13}) imply
\begin{eqnarray}
\label{2.14}
\exp \{ \beta \sum_{s_{i}^{1} \in P(G)}
E(s_{i}^{1})(- 1)^{\partial^{\star} \sigma (s_{i}^{1})}\}
= \nonumber \\
\left( \prod_{s^{1}_{i} \in P(G)} \cosh \beta E(s^{1}_{i})\right)
\sum_{\xi^{1} \in C^{1}(P(G),Z^{add}_{2})}
(- 1)^{\langle \sigma ,\partial \xi^{1} \rangle }
\prod_{s^{1}_{i} \in P(G)}
(\tanh \beta E(s^{1}_{i}))^{\frac{1}{2} (1 - (- 1)^{\xi^{1} (s^{1}_{i})})},
\end{eqnarray}
\begin{eqnarray}
\label{2.15}
\exp \{ \beta (\# (VG))^{- 1/2}H\sum_{s_{i}^{0} \in P(G)}
(- 1)^{\sigma (s_{i}^{0})}\} =
(\cosh \beta (\# (VG))^{- 1/2}H)^{\# (VG)} \times \nonumber \\
\sum_{\xi^{0} \in C^{0}(P(G),Z_{2}^{add})}
(- 1)^{\langle \sigma ,\xi^{0} \rangle}
(\tanh \beta (\# (VG))^{- 1/2}H)^{\sum_{s_{i}^{0} \in P(G)}
\frac{1}{2} (1 - (- 1)^{\xi^{0} (s_{i}^{0})})}
\end{eqnarray}
The substitution of the equalities (\ref{2.14}), (\ref{2.15}) and the relation
\begin{equation}
\label{2.16}
\sum_{\xi = 0,1} (- 1)^{\xi \epsilon} =
\left\{ {2, \hskip 1cm \epsilon = 0,} \atop
{0, \hskip 1cm \epsilon = 1,} \right.
\end{equation}
into the definition (\ref{2.8}) gives
\begin{eqnarray}
\label{2.17} Z_{G}((\# (VG))^{- 1/2}H) = 2^{\# (VG)}(\cosh \beta (\# (VG))^{-
1/2}H)^{\# (VG)} \left( \prod_{s^{1}_{i} \in P(G)} \cosh \beta E(s^{1}_{i})\right)
\times  \nonumber \\ Z_{r,G}((\# (VG))^{- 1/2}H),
\end{eqnarray}
\begin{eqnarray}
\label{2.18} Z_{r,G}((\# (VG))^{- 1/2}H) = \sum_{\xi^{1} \in C^{1}(P(G),Z^{add}_{2})}
\left( \prod_{s^{1}_{i} \in P(G)} (\tanh \beta E(s^{1}_{i}))^{\frac{1}{2} (1 - (-
1)^{\xi^{1} (s^{1}_{i})})}
\right) \times \nonumber \\
(\tanh \beta (\# (VG))^{- 1/2}H)^{\sum_{s_{i}^{0} \in P(G)}
\frac{1}{2} (1 - (- 1)^{\partial \xi^{1} (s_{i}^{0})})}.
\end{eqnarray}

We denote the sum (\ref{2.8}) for $H = 0$ by $Z_{G}$, the sum (\ref{2.9})
for $H = 0$ by $W_{G}(\chi )$ and the sum (\ref{2.18}) for $H = 0$ by
$Z_{r,G}$. The relation (\ref{2.18}) implies
\begin{equation}
\label{2.19}
Z_{r,G} = \sum_{\xi^{1} \in Z_{1}(P(G),Z^{add}_{2})}
\prod_{s^{1}_{i} \in P(G)}
(\tanh \beta E(s^{1}_{i}))^{\frac{1}{2} (1 - (- 1)^{\xi^{1} (s^{1}_{i})})}.
\end{equation}
Due to Proposition 3.1 from the paper \cite{7}
\begin{equation}
\label{2.20}
W_{G}(\chi ) = (Z_{r,G})^{- 1}
\sum_{{\xi^{1} \in C^{1}(P(G),Z^{add}_{2}),} \atop
{\partial \xi^{1} \, = \, \chi }}
\prod_{s^{1}_{i} \in P(G)}
(\tanh \beta E(s^{1}_{i}))^{\frac{1}{2} (1 - (- 1)^{\xi^{1} (s^{1}_{i})})}
\end{equation}
The substitution of the equality (\ref{2.20}) into the equality (\ref{2.18})
gives
\begin{eqnarray}
\label{2.21} (Z_{r,G})^{-1}Z_{r,G}((\# (VG))^{- 1/2}H) = \sum_{\xi^{0} \in
B_{0}(P(G),Z_{2}^{add})} W_{G}(\xi^{0} ) \times \nonumber \\ (\tanh \beta (\# (VG))^{-
1/2}H)^{\sum_{s_{i}^{0} \in P(G)} \frac{1}{2} (1 - (- 1)^{\xi^{0} (s_{i}^{0})})}.
\end{eqnarray}

The sum (\ref{2.19}) is independent of $H$. Hence the definition (\ref{2.12})
and the relation (\ref{2.17}) imply
\begin{eqnarray}
\label{2.22}
m_{G}(H) = \beta^{- 1} \frac{\partial}{\partial H}
\ln ((Z_{r,G})^{- 1}Z_{G}((\# (VG))^{- 1/2}H) ) = \nonumber \\
(\# (VG))^{1/2}\tanh \beta (\# (VG))^{- 1/2}H + \beta^{- 1} \frac{\partial}{\partial H}
\ln ((Z_{r,G})^{- 1}Z_{r,G}((\# (VG))^{- 1/2}H)),
\end{eqnarray}
\begin{equation}
\label{2.23} \lim_{G \rightarrow Z^{\times 2}} m_{G}(H) = \beta H + \lim_{G \rightarrow
Z^{\times 2}} \beta^{- 1} \frac{\partial}{\partial H} \ln ((Z_{r,G})^{- 1}Z_{r,G}((\#
(VG))^{- 1/2}H)).
\end{equation}

Let us consider the function (\ref{2.21}). The cochain
$\xi^{0} \in B_{0}(P(G),{\bf Z}_{2}^{add})$ takes the value $1$ only
at the ends of the broken lines. Any broken line has two ends. Hence
the cochain $\xi^{0}$ takes the value $1$ at the even number of vertices.
Conversely, since the graph $G(M_{1}^{\prime},M_{2}^{\prime};M_{1},M_{2})$
is arcwise connected, any even number of vertices may be pairwise connected
by the broken lines. Let the cochain $\xi^{0}$ equal $1$ on the even number
of vertices. Then it is the boundary of the cochain $\xi^{1}$ which is
equal to $1$ on the edges of these broken lines. Thus the group
$B_{0}(P(G),{\bf Z}_{2}^{add})$ consists of the cochains
\begin{equation}
\label{2.25}
\xi^{0} = \sum_{j\, =\, 1}^{2k} \delta^{0} [s_{i_{j}}^{0}],
\end{equation}
\begin{equation}
\label{2.26}
\delta^{0} [s_{i}^{0}](s_{j}^{0}) = \delta_{ij}.
\end{equation}
Therefore the relation (\ref{2.21}) may be rewritten as
\begin{eqnarray}
\label{2.27} (Z_{r,G})^{- 1}Z_{r,G}((\# (VG))^{- 1/2}H) = 1 + \sum_{k\, =\,
1}^{[\frac{1}{2} \# (VG)]} (\tanh \beta (\# (VG))^{- 1/2}H)^{2k} \times \nonumber \\
\frac{1}{(2k)!} \sum_{{s_{i_{1}}^{0}\, \neq\, \cdots \neq \, s_{i_{2k}}^{0},}\atop
{s_{i_{j}}^{0}\, \in \, P(G)}} W_{G}(\sum_{j\, =\, 1}^{2k} \delta^{0} [s_{i_{j}}^{0}]).
\end{eqnarray}
since a permutation of the vertices $s_{i_{j}}^{0}$ does not change the cochain
$\sum_{j\, =\, 1}^{2k} \delta^{0} [s_{i_{j}}^{0}]$.

The cochain $\xi^{1} \in C^{1}(P(G),{\bf Z}_{2}^{add})$ is called the
non -- oriented path connecting the vertices $s_{i}^{0}$ and $s_{j}^{0}$
if
\begin{equation}
\label{2.28}
\partial \xi^{0} = \delta^{0} [s_{i}^{0}] + \delta^{0} [s_{j}^{0}]
\end{equation}
and any edge of the path is connected at any vertex to at most one other
edge of the path. Any cochain from the group $C^{1}(P(G),{\bf Z}_{2}^{add})$
may be represented as a sum of the non -- oriented paths. These non --
oriented paths have no common edges. If we suppose that these non --
oriented paths self -- intersect transversally and intersect each other
transversally, then this representation is unique. In order to establish
this representation it is sufficient to consider two cases: the cochain
$\xi^{1} \in C^{1}(P(G),{\bf Z}_{2}^{add})$ is equal to $1$ on three or
four edges incident to one vertex. Let the cochain be equal to $1$ on
the three edges incident to one vertex. Two edges of these three edges are
vertical or horizontal. We connect these two edges in one non -- oriented
path. Let the cochain be equal to $1$ on the four edges incident to one
vertex. Two edges of these four edges are horizontal and last two edges are
vertical. We connect the horizontal edges in one non -- oriented path and
connect the vertical edges in another (in general) non -- oriented path. Thus
the set of edges on which the cochain is equal to $1$ is divided into the
set of the non -- oriented paths. By construction these non -- oriented
paths self -- intersect transversally and intersect each other transversally.
Therefore
\begin{equation}
\label{2.29}
\xi^{1} = \sum_{i\, =\, 1}^{m} \xi_{i}^{1} + \sum_{j\, =\, 1}^{n} \eta_{j}^{1}
\end{equation}
where $\xi_{i}^{1}$ is a non -- oriented path connecting two vertices
and $\eta_{j}^{1}$ is a non -- oriented closed path. The non -- oriented
paths $\xi_{i}^{1}, \eta_{j}^{1}$ self -- intersect transversally and
intersect each other transversally.

Let us denote $j_{1}(\xi^{1} )(j_{2}(\xi^{1} ))$ the cochain which is
equal to $1$ on the horizontal (vertical) edges incident to the vertices
incident to the horizontal (vertical) edges on which the cochain $\xi^{1}$
is equal to $1$. We define
\begin{equation}
\label{2.30}
j(\xi^{1} ) = j_{1}(\xi^{1} ) + j_{2}(\xi^{1} ).
\end{equation}
The support $||\xi^{1} ||$ is the set of all edges of the graph
$G(M_{1}^{\prime},M_{2}^{\prime};M_{1},M_{2})$ at which a cochain
$\xi^{1} $ takes the value $1$. The non -- oriented paths
$\xi_{i}^{1}, \eta_{j}^{1}$ in the equality (\ref{2.29}) intersect each
other transversally. This condition may be written as
\begin{equation}
\label{2.31}
||j(\xi_{i}^{1} )||\cap ||\eta_{j}^{1} || = \emptyset.
\end{equation}
By making use of the equality (\ref{2.29}) we rewrite the relation
(\ref{2.20}) in the following form
\begin{equation}
\label{2.32}
W_{G}(\chi ) = \sum_{\sum_{i} \partial \xi_{i}^{1} \, =\, \chi }
\left( \prod_{i\, =\, 1}^{m(\chi )} (\tanh \beta E)^{\xi_{i}^{1}}\right)
(Z_{r,G})^{- 1}Z_{r,G^{\prime}},
\end{equation}
\begin{equation}
\label{2.33}
(\tanh \beta E)^{\xi^{1} } = \prod_{s_{i}^{1}\, \in \, P(G)}
(\tanh \beta E(s_{i}^{1}))^{\frac{1}{2} (1 - (- 1)^{\xi^{1} (s_{i}^{1})})},
\end{equation}
\begin{equation}
\label{2.34}
G^{\prime} = G\setminus \cup_{i\, =\, 1}^{m(\chi )} ||j(\xi_{i}^{1} )||.
\end{equation}
where in the equality (\ref{2.32}) the summing runs over the set of the
non -- oriented paths $\xi_{i}^{1}$ connecting pairwise the vertices on
which the cochain $\chi$ is equal to $1$. These non -- oriented paths
$\xi_{i}^{1}$ have no common edges. These non -- oriented paths
self -- intersect transversally and intersect each other transversally.
If the number of vertices on which the cochain $\chi$ is equal to $1$
is equal to $2m(\chi )$, then the number of these non -- oriented paths
$\xi_{i}^{1}$ is equal to $m(\chi )$. The graph (\ref{2.34}) is obtained
from the graph $G(M_{1}^{\prime},M_{2}^{\prime};M_{1},M_{2})$ by deleting
all edges from the supports $||j(\xi_{i}^{1} )||, i = 1,...,m(\chi )$.
For any graph $G^{\prime}$ embedded in the rectangular lattice
$G(M_{1}^{\prime},M_{2}^{\prime};M_{1},M_{2})$ it is possible to define
the cell complex $P(G^{\prime})$. Then the reduced partition function
$Z_{r,G^{\prime}}$ is given by the relation (\ref{2.19}).

We write the edge $s_{i}^{1}$ of the lattice
$G(M^{\prime}_{1},M^{\prime}_{2};M_{1},M_{2})$ as the pair
$\{ {\bf p},{\bf e}\} $ where ${\bf p}$ is the edge initial vertex and
the unit vector ${\bf e}$ is the edge direction. The unit vector
${\bf e}$ is one of the four vectors: $(\pm 1,0)$, $(0,\pm 1)$.
Since the edge is non -- oriented, then
$\{ {\bf p} + {\bf e}, - {\bf e}\} = \{ {\bf p},{\bf e}\} $. The intitial
vertex ${\bf p}$ and the final vertex ${\bf p} + {\bf e}$ of the edge
$\{ {\bf p},{\bf e}\} $ are the vertices of the graph
$G(M^{\prime}_{1},M^{\prime}_{2};M_{1},M_{2})$.
Hence the components $p^{i}, i = 1,2$, are the integers and
$M^{\prime}_{i} \leq p^{i} \leq M_{i}$,
$M^{\prime}_{i} \leq p^{i} + e^{i} \leq M_{i}$, $i = 1,2$.

We write the oriented edge of the lattice
$G(M^{\prime}_{1},M^{\prime}_{2};M_{1},M_{2})$ as the pair
$({\bf p},{\bf e})$ where ${\bf p}$ is the oriented edge initial vertex and
the unit vector ${\bf e}$ is the oriented edge direction. Let
$\# ({\bf E}G)$ be the total number of the oriented edges of the graph
$G(M^{\prime}_{1},M^{\prime}_{2};M_{1},M_{2})$. We define
$\# ({\bf E}G) \times \# ({\bf E}G)$ -- matrix
\begin{equation}
\label{2.35}
T_{(p_{1},e_{1}),(p_{2},e_{2})} =
\left\{ {\exp\{\frac{i}{2} \hat{({\bf e}_{1},{\bf e}_{2})} \}
\tanh \beta E(\{ {\bf p}_{1},{\bf e}_{1}\}), \hskip 0,5cm
{\bf p}_{2}\, =\, {\bf p}_{1}\, +\, {\bf e}_{1},\,
{\bf p}_{1}\, \neq \, {\bf p}_{2}\, +\, {\bf e}_{2},} \atop
{0, \hskip 2,5cm otherwise} \right.
\end{equation}
where $\hat{({\bf e}_{1},{\bf e}_{2})}$ is the radian measure of the
angle between the direction of the unit vector ${\bf e}_{1}$ and the
direction of the unit vector ${\bf e}_{2}$. Due to the paper \cite{8}
\begin{equation}
\label{2.36}
(Z_{r,G})^{2} = \det (I - T)
\end{equation}
where the reduced partition function $Z_{r,G}$ is given by the relation
(\ref{2.19}) and $I$ is the identity $\# ({\bf E}G) \times \# ({\bf E}G)$ --
matrix. The formula (\ref{2.36}) is obtained for an arbitrary graph $G$
embedded in a lattice on the plane. We suppose that a graph $G$ is
embedded in the lattice $G(M^{\prime}_{1},M^{\prime}_{2};M_{1},M_{2})$.
Let the estimate
\begin{equation}
\label{2.37}
|\tanh \beta E(\{ {\bf p},{\bf e}\} )| < 1/3
\end{equation}
be fulfilled for any edge of the graph $G$. Then
\begin{equation}
\label{2.38}
\det (I - T) = \exp \{ - \sum_{k\, =\, 1}^{\infty} k^{- 1}\hbox{tr} T^{k}\}.
\end{equation}

A closed path is a sequence of the oriented edges
$C = (({\bf p}_{1},{\bf e}_{1}),...,({\bf p}_{k},{\bf e}_{k}))$ such that
\begin{equation}
\label{2.39}
{\bf p}_{i + 1} = {\bf p}_{i} + {\bf e}_{i}, \, i = 1,...,k - 1, \,
{\bf p}_{1} = {\bf p}_{k} + {\bf e}_{k}.
\end{equation}
The number $|C| = k$ is called the length of the closed path
$C = (({\bf p}_{1},{\bf e}_{1}),...,({\bf p}_{k},{\bf e}_{k}))$.
The support $||C||$ is the set of all edges
$\{ {\bf p},{\bf e}\} $ such that the oriented
edge $({\bf p},{\bf e})$ is included into the path $C$ or the oriented
edge $({\bf p} + {\bf e}, - {\bf e})$ is included into the path $C$.
The closed path $(({\bf p}_{1},{\bf e}_{1}),...,({\bf p}_{k},{\bf e}_{k}))$
is called reduced if it satisfies the following condition
\begin{equation}
\label{2.40}
{\bf p}_{i + 1} = {\bf p}_{i} + {\bf e}_{i}, \,
{\bf p}_{i + 1} + {\bf e}_{i + 1} \neq {\bf p}_{i}, \, i = 1,...,k - 1, \,
{\bf p}_{1} = {\bf p}_{k} + {\bf e}_{k}, \,
{\bf p}_{1} + {\bf e}_{1} \neq {\bf p}_{k}.
\end{equation}
The set of all reduced closed paths on the graph $G$ is denoted by $RC(G)$.
With any reduced closed path
$C = (({\bf p}_{1},{\bf e}_{1}),...,({\bf p}_{k},{\bf e}_{k}))$ on the
graph $G$ there corresponds the total angle through which the tangent vector
of the path $C$ turns along the path $C$
\begin{equation}
\label{2.41}
\phi (C) = \hat{({\bf e}_{1},{\bf e}_{2})} + \hat{({\bf e}_{2},{\bf e}_{3})}
+ \cdots + \hat{({\bf e}_{k - 1},{\bf e}_{k})} +
\hat{({\bf e}_{k},{\bf e}_{1})}.
\end{equation}
Due to the paper \cite{7}
\begin{equation}
\label{2.42}
\hbox{tr} T^{k} = \sum_{{C\, \in \, RC(G),} \atop {|C|\, =\, k}}
\exp \{ \frac{i}{2} \phi (C)\} (\tanh \beta E)^{C},
\end{equation}
\begin{equation}
\label{2.43}
(\tanh \beta E)^{C} = \prod_{(p,e)\, \in \, C}
\tanh \beta E(\{ {\bf p},{\bf e}\} ).
\end{equation}
The substitution of the equality (\ref{2.42}) into the equalities
(\ref{2.36}), (\ref{2.38}) gives
\begin{equation}
\label{2.44}
(Z_{r,G})^{2} = \exp \{ - \sum_{C\, \in \, RC(G)} |C|^{- 1}
\exp \{ \frac{i}{2} \phi (C)\} (\tanh \beta E)^{C}\}.
\end{equation}
In view of the definition the angle $\phi (C) = 2\pi k$ where $k$ is
an integer. Hence the number $\exp \{ \frac{i}{2} \phi (C)\}$ is real.
By the definitions (\ref{2.8}), (\ref{2.17}) the number $Z_{r,G}$ is
positive. Thus the equality (\ref{2.44}) implies
\begin{equation}
\label{2.45}
Z_{r,G} = \exp \{ - \frac{1}{2} \sum_{C\, \in \, RC(G)} |C|^{- 1}
\exp \{ \frac{i}{2} \phi (C)\} (\tanh \beta E)^{C}\}.
\end{equation}

By the definition a reduced closed path does not contain the oppositely
oriented edges $({\bf p},{\bf e}), ({\bf p} + {\bf e}, - {\bf e})$ if they
are subsequent or if they are the first and the last edges of the closed
path. A closed path is called completely reduced if it does not contain
the oppositely oriented edges
$({\bf p},{\bf e}), ({\bf p} + {\bf e}, - {\bf e})$ at any place. The set
of all completely reduced closed paths on the graph $G$ is denoted by
$CRC(G)$.

{\bf Theorem 2.1.} {\it Let a graph} $G$ {\it be embedded in the rectagular lattice}
$G(M^{\prime}_{1},M^{\prime}_{2};M_{1},M_{2})$ {\it on the plane. Let with any reduced
closed path} $C$ {\it on the graph} $G$ {\it there correspond the number} $\phi (C)$
{\it given by the relation} (\ref{2.41}) {\it and the number} $(\tanh \beta E)^{C}$
{\it given by the relation} (\ref{2.43}). {\it If the estimate} (\ref{2.37}) {\it is
valid, then}
\begin{eqnarray}
\label{2.46}
\sum_{C\, \in \, RC(G)} |C|^{- 1}
\exp \{ \frac{i}{2} \phi (C)\} (\tanh \beta E)^{C} = \nonumber \\
\sum_{C\, \in \, CRC(G)} |C|^{- 1}
\exp \{ \frac{i}{2} \phi (C)\} (\tanh \beta E)^{C}.
\end{eqnarray}
{\it In the left hand side of the equality} (\ref{2.46}) {\it the sum
extends over the set} $RC(G)$ {\it of all the reduced closed paths on
the graph} $G$ {\it and in the right hand side of the equality} (\ref{2.46})
{\it the sum extends over the set} $CRC(G)$ {\it of all the completely
reduced closed paths on the graph} $G$.

{\bf Proof.} Let the reduced closed path $C = (({\bf p}_{1},{\bf e}_{1}),...,({\bf
p}_{p},{\bf e}_{p}), ({\bf p},{\bf e}), ({\bf p}_{p + 1},{\bf e}_{p + 1}),...,$

\noindent $({\bf p}_{p + q},{\bf e}_{p + q}),
({\bf p} + {\bf e}, - {\bf e}),({\bf p}_{p + q + 1},{\bf e}_{p + q + 1}),
...,({\bf p}_{p + q + r},{\bf e}_{p + q + r}))$ contain the oppositely
oriented edges $({\bf p},{\bf e})$ and $({\bf p} + {\bf e}, - {\bf e})$.
Then the closed path $C^{\prime} = (({\bf p}_{1},{\bf e}_{1}),...,
({\bf p}_{p},{\bf e}_{p}),({\bf p},{\bf e}),
({\bf p}_{p + q} + {\bf e}_{p + q}, - {\bf e}_{p + q}),...,
({\bf p}_{p + 1} + {\bf e}_{p + 1}, - {\bf e}_{p + 1}),
({\bf p} + {\bf e}, - {\bf e}),({\bf p}_{p + q + 1},{\bf e}_{p + q + 1}),
...,({\bf p}_{p + q + r},{\bf e}_{p + q + r}))$ is also reduced. The
path length definition and the definition (\ref{2.43}) imply
\begin{equation}
\label{2.47}
|C| = |C^{\prime}|, \, \, (\tanh \beta E)^{C} =
(\tanh \beta E)^{C^{\prime}}.
\end{equation}
By using the definition (\ref{2.41}) we get
\begin{equation}
\label{2.48}
\phi (C) = \phi_{1} + \phi_{2} + \phi_{3},
\end{equation}
\begin{equation}
\label{2.49}
\phi_{1} = \sum_{i\, =\, 1}^{p\, -\, 1} \hat{({\bf e}_{i},{\bf e}_{i + 1})}
+ \hat{({\bf e}_{p},{\bf e})},
\end{equation}
\begin{equation}
\label{2.50}
\phi_{2} = \hat{({\bf e},{\bf e}_{p + 1})} +
\sum_{i\, =\, p\, +\, 1}^{p\, +\, q\,  -\, 1}
\hat{({\bf e}_{i},{\bf e}_{i + 1})}
+ \hat{({\bf e}_{p + q}, - {\bf e})},
\end{equation}
\begin{equation}
\label{2.51}
\phi_{3} = \hat{(- {\bf e},{\bf e}_{p + q + 1})} +
\sum_{i\, =\, p\, +\, q\, +\, 1}^{p\, +\, q\, +\, r\, -\, 1}
\hat{({\bf e}_{i},{\bf e}_{i + 1})}
+ \hat{({\bf e}_{p + q + r},{\bf e}_{1})}.
\end{equation}
It is easy to verify the relation
\begin{equation}
\label{2.52}
\hat{({\bf e}_{1},{\bf e}_{2})} =
- \hat{( - {\bf e}_{2}, - {\bf e}_{1})}.
\end{equation}
The definition (\ref{2.41}) and the relation (\ref{2.52}) imply
\begin{equation}
\label{2.53}
\phi (C^{\prime}) = \phi_{1} - \phi_{2} + \phi_{3}.
\end{equation}
Due to the definition (\ref{2.50}) $\phi_{2} = (2k + 1)\pi $ where
$k$ is an integer. Hence $\exp \{ - \frac{i}{2} \phi_{2} \} =
- \exp \{ \frac{i}{2} \phi_{2} \} $. In view of the relations (\ref{2.48}),
(\ref{2.53})
\begin{equation}
\label{2.54}
\exp \{ \frac{i}{2} \phi (C)\} = - \exp \{ \frac{i}{2} \phi (C^{\prime})\}.
\end{equation}
Due to the relations (\ref{2.47}), (\ref{2.54}) all terms
$|C|^{- 1}\exp \{ \frac{i}{2} \phi (C)\} (\tanh \beta E)^{C}$ in the
left hand side sum (\ref{2.46}) corresponding with the reduced closed
paths $C$ containing the oppositely oriented edges $({\bf p},{\bf e})$
and $({\bf p} + {\bf e}, - {\bf e})$ cancel each other. The theorem is
proved.

Since the relations (\ref{2.45}), (\ref{2.46}) are valid for any graph
embedded in the rectangular lattice
$G(M^{\prime}_{1},M^{\prime}_{2};M_{1},M_{2})$, then
\begin{equation}
\label{2.55}
(Z_{r,G})^{- 1}Z_{r,G^{\prime}} =
\exp \{ \frac{1}{2} \sum_{{C\, \in \, CRC(G),}\atop
{||C||\, \cap \, \cup_{i} \, ||j(\xi_{i}^{1})|| \, \neq \, \emptyset }}
|C|^{- 1}\exp \{ \frac{i}{2} \phi (C)\} (\tanh \beta E)^{C}\}
\end{equation}
where the graph $G^{\prime}$ is given by the relation (\ref{2.34})
and the estimate (\ref{2.37}) is valid.

The substitution of the equality (\ref{2.55}) into the equality (\ref{2.32})
gives
\begin{eqnarray}
\label{2.56}
W_{G}(\chi ) = \sum_{\sum_{i} \partial \xi_{i}^{1} \, =\, \chi }
\left( \prod_{i\, =\, 1}^{m(\chi )} (\tanh \beta E)^{\xi_{i}^{1}}\right)
\times \nonumber \\
\exp \{ \frac{1}{2} \sum_{{C\, \in \, CRC(G),}\atop
{||C||\, \cap \, \cup_{i} \, ||j(\xi_{i}^{1})|| \, \neq \, \emptyset }}
|C|^{- 1}\exp \{ \frac{i}{2} \phi (C)\} (\tanh \beta E)^{C}\}
\end{eqnarray}
where the summing runs over the set of the non -- oriented paths
$\xi_{i}^{1}$ connecting pairwise the vertices on which the cochain
$\chi $ is equal to $1$. These non -- oriented paths have no common
edges. These non -- oriented paths self -- intersect transversally and
intersect each other transversally. If the number of vertices on which
the cochain $\chi $ is equal to $1$ is equal to $2m(\chi )$, then the
number of these non -- oriented paths $\xi_{i}^{1} $ is equal to $m(\chi )$.

\section{Thermodynamic limit}
\setcounter{equation}{0}

\noindent In this section we study the limit of the formulae (\ref{2.27}),
(\ref{2.56}) when the graph

\noindent $G(M^{\prime}_{1},M^{\prime}_{2};M_{1},M_{2})$
tends to the infinite graph ${\bf Z}^{\times 2}$, i.e.
$M_{i}^{\prime} \rightarrow - \infty $, $M_{i} \rightarrow \infty $,
$i = 1,2$.

Let us construct on the graph ${\bf Z}^{\times 2}$ a reduced closed path
with the fixed initial vertex. For the first oriented edge we have $4$
possibilities. For any other oriented edge the number of possibilities is
not more than $3$. Thus the total number of reduced closed paths of the
length $l$ starting at the fixed vertex is not more than $4\cdot 3^{l - 1}$.
Hence the series
\begin{equation}
\label{3.1}
\sum_{{C\, \in \, CRC(Z^{\times 2}),}\atop
{||C||\, \cap \, \cup_{i} \, ||j(\xi_{i}^{1})|| \, \neq \, \emptyset }}
|C|^{- 1}\exp \{ \frac{i}{2} \phi (C)\} (\tanh \beta E)^{C}
\end{equation}
is absolutely convergent if the estimate (\ref{2.37}) is valid.

We suppose that all interaction energies $E(\{ {\bf p},{\bf e}\} )$
attached to the vertical (horizontal) edges $\{ {\bf p},{\bf e}\}$
have the same sign. Any closed path $C$ on the lattice ${\bf Z}^{\times 2}$
has an even number of the verically directed edges and it has an even
number of the horizontally directed edges. Therefore
\begin{equation}
\label{3.2}
(\tanh \beta E)^{C} \geq 0.
\end{equation}
Any cycle $\xi^{1} \in Z_{1}(P(G),{\bf Z}_{2}^{add})$ for a graph
$G$ embedded in the lattice ${\bf Z}^{\times 2}$ may be represented
as the set of closed non -- oriented paths on the lattice
${\bf Z}^{\times 2}$. These non -- oriented paths self -- intersect
transversally and intersect each other transversally. Hence the
inequality (\ref{3.2}) implies
\begin{equation}
\label{3.3}
(\tanh \beta E)^{\xi^{1}} \geq 0.
\end{equation}
It follows from the equality (\ref{2.19}) and the inequality (\ref{3.3})
that
\begin{equation}
\label{3.4}
(Z_{r,G})^{- 1}Z_{r,G^{\prime}} \leq 1
\end{equation}
where the graph $G^{\prime}$ is given by the relation (\ref{2.34}). By
making use of the inequality (\ref{3.4}) and the equality (\ref{2.32})
for $G \rightarrow {\bf Z}^{\times 2}$ we get the estimate
\begin{equation}
\label{3.5}
\mid W_{Z^{\times 2}}(\chi )\mid \leq
\sum_{\sum_{i} \partial \xi_{i}^{1} \, =\, \chi }
\prod_{i\, =\, 1}^{m(\chi )} |(\tanh \beta E)^{\xi_{i}^{1}}|
\end{equation}
where the summing runs over the set of the non -- oriented paths
$\xi_{i}^{1}$ connecting pairwise the vertices on which the cochain
$\chi $ is equal to $1$. These non -- oriented paths have no common
edges. These non -- oriented paths self -- intersect transversally and
intersect each other transversally. If the number of vertices on which
the cochain $\chi $ is equal to $1$ is equal to $2m(\chi )$, then the
number of these non -- oriented paths $\xi_{i}^{1} $ is equal to $m(\chi )$.
For the fixed cochain $\chi $ the series (\ref{3.5}) is absolutely
convergent if the estimate (\ref{2.37}) is valid. Therefore for
$G \rightarrow {\bf Z}^{\times 2}$ the sum (\ref{2.56}) for the fixed
cochain $\chi $ tends to the absolutely convergent series
\begin{eqnarray}
\label{3.6}
W_{Z^{\times 2}}(\chi ) = \sum_{\sum_{i} \partial \xi_{i}^{1} \, =\, \chi }
\left( \prod_{i\, =\, 1}^{m(\chi )} (\tanh \beta E)^{\xi_{i}^{1}}\right)
\times \nonumber \\
\exp \{ \frac{1}{2} \sum_{{C\, \in \, CRC(Z^{\times 2}),}\atop
{||C||\, \cap \, \cup_{i} \, ||j(\xi_{i}^{1})|| \, \neq \, \emptyset }}
|C|^{- 1}\exp \{ \frac{i}{2} \phi (C)\} (\tanh \beta E)^{C}\}.
\end{eqnarray}

The cells $s_{i}^{0}$ are the vertices of the graph
$G(M^{\prime}_{1},M^{\prime}_{2};M_{1},M_{2})$ or the vectors
${\bf p} \in {\bf R}^{2}$ with the integer components and
$M^{\prime}_{i} \leq p^{i} \leq M_{i}, i = 1,2$. We define the cochain
$\delta^{0} [{\bf p}]$ similarly to the definition (\ref{2.26})
\begin{equation}
\label{3.7}
\delta^{0} [{\bf p}]({\bf q}) = \delta_{p^{1},q^{1}} \delta_{p^{2},q^{2}}.
\end{equation}

{\bf Proposition 3.1.} {\it Let the interaction energy} $E(\{ {\bf p},{\bf e}\} )$ {\it
depend only on the direction (horizontal or vertical) of the edge} $\{ {\bf p},{\bf
e}\} $. {\it Let the estimate} (\ref{2.37}) {\it be valid. Then}
\begin{equation}
\label{3.9} \lim_{G \rightarrow Z^{\times 2}} (Z_{r,G})^{- 1}Z_{r,G}((\# (VG))^{-
1/2}H) = \exp \{ \frac{(\beta H)^{2}}{2} \sum_{p\, \in \, Z^{\times 2},\, p\, \neq \,
0} W_{Z^{\times 2}}(\delta^{0} [{\bf 0}] + \delta^{0} [{\bf p}])\}
\end{equation}
{\it where the correlation function} $W_{Z^{\times 2}}(\chi )$ {\it is given by the
relation} (\ref{3.6}).

{\bf Proof.} We divide the numbers $1,...,2l$ into $l$ pairs
$i_{1},j_{1};...;i_{l},j_{l}$ where $1 = i_{1} < \cdots < i_{l}$ and $i_{1} <
j_{1},...,i_{l} < j_{l}$. The total number of such subdivisions is equal to
$(2l)!(l!)^{- 1}2^{- l}$. In the relation (\ref{2.32}) for the correlation function
$W_{G}(\sum_{i\, =\, 1}^{2l} \delta^{0} [{\bf p}_{i}])$ the summing runs over the set
of such subdivisions of the numbers $1,...,2l$ and over the set of non -- oriented
paths $\xi_{i_{1}}^{1}, ...,\xi_{i_{l}}^{1}$ connecting the vertices ${\bf p}_{i_{s}}$
and ${\bf p}_{j_{s}}, s = 1,...,l$. By making the changes of summing variables we get
\begin{eqnarray}
\label{3.11}
\sum_{p_{1}\, \neq \, \cdots \, \neq \, p_{2l},\, p_{i}\, \in \, P(G)}
W_{G}(\sum_{i\, =\, 1}^{2l} \delta^{0} [{\bf p}_{i}]) = \nonumber \\
2^{- l}\frac{(2l)!}{l!}
\sum_{p_{1}\, \neq \, \cdots \, \neq \, p_{2l},\, p_{i}\, \in \, P(G)}
\sum_{{\partial \xi_{i}^{1} \, =\, \delta^{0} [p_{2i - 1}]\, +\,
\delta^{0} [p_{2i}],}\atop {i = 1,...,l}}
\left( \prod_{i\, =\, 1}^{l} (\tanh \beta E)^{\xi_{i}^{1}}\right)
(Z_{r,G})^{- 1}Z_{r,G^{\prime}}.
\end{eqnarray}
The equality (\ref{3.11}) and the estimate (\ref{3.4}) imply
\begin{eqnarray}
\label{3.12}
\mid \sum_{p_{1}\, \neq \, \cdots \, \neq \, p_{2l},\, p_{i}\, \in \, P(G)}
W_{G}(\sum_{i\, =\, 1}^{2l} \delta^{0} [{\bf p}_{i}])\mid \leq \nonumber \\
2^{- l}\frac{(2l)!}{l!}
\left( \sum_{p_{1}\, \neq \, p_{2},\, p_{i}\, \in \, P(G)}
\sum_{\partial \xi^{1} \, =\, \delta^{0} [p_{1}]\, +\, \delta^{0} [p_{2}]}
\mid (\tanh \beta E)^{\xi^{1}}\mid \right)^{l}.
\end{eqnarray}
In the right hand side of the inequality (\ref{3.12}) instead of the
non -- oriented paths $\xi_{1}^{1}, ...,\xi_{l}^{1}$ which self --
intersect and intersect each other transversally we consider the non --
oriented paths $\xi_{1}^{1}, ...,\xi_{l}^{1}$ which self -- intersect
transversally. We may also consider such non -- oriented  paths
$\xi_{1}^{1}, ...,\xi_{l}^{1}$ on the lattice ${\bf Z}^{\times 2}$.
Since the interaction energy $E(\{ {\bf p},{\bf e}\} )$ depends only on
the direction (horizontal or vertical) of the edge $\{ {\bf p},{\bf e}\}$,
then in view of the definition (\ref{2.33})
\begin{eqnarray}
\label{3.13}
\sum_{p_{1}\, \neq \, p_{2},\, p_{i}\, \in \, P(G)}
\sum_{\partial \xi^{1} \, =\, \delta^{0} [p_{1}]\, +\, \delta^{0} [p_{2}]}
\mid (\tanh \beta E)^{\xi^{1}}\mid = \nonumber \\
\sum_{q\, \in \, Z^{\times 2},\, q\, \neq \, 0}
\sum_{\partial \xi^{1} \, =\, \delta^{0} [0]\, +\, \delta^{0} [q]}
\mid (\tanh \beta E)^{\xi^{1}}\mid
\sum_{p_{1}, p_{2}\, \in \, P(G),\, p_{2}\, -\, p_{1}\, =\, q} 1.
\end{eqnarray}
It is easy to verify that
\begin{equation}
\label{3.14}
\sum_{p_{1}, p_{2}\, \in \, P(G),\, p_{2}\, -\, p_{1}\, =\, q} 1 \leq
\# (VG).
\end{equation}
It follows from the estimates (\ref{3.12}), (\ref{3.14}) and from the
equality (\ref{3.13}) that
\begin{eqnarray}
\label{3.15}
\mid \sum_{p_{1}\, \neq \, \cdots \, \neq \, p_{2l},\, p_{i}\, \in \, P(G)}
W_{G}(\sum_{i\, =\, 1}^{2l} \delta^{0} [{\bf p}_{i}])\mid \leq \nonumber \\
2^{- l}\frac{(2l)!}{l!} (\# (VG))^{l}
\left( \sum_{q\, \in \, Z^{\times 2},\, q\, \neq \, 0}
\sum_{\partial \xi^{1} \, =\, \delta^{0} [0]\, +\, \delta^{0} [q]}
\mid (\tanh \beta E)^{\xi^{1}}\mid \right)^{l}
\end{eqnarray}
where the second summing in the right hand side of the inequality
(\ref{3.15}) runs the non -- oriented paths $\xi^{1}$ on the lattice
${\bf Z}^{\times 2}$ connecting the vertices $0$ and ${\bf q}$.
These non -- oriented paths self -- intersect transversally. In view
of the estimate (\ref{2.37}) the series
\begin{equation}
\label{3.16}
\sum_{q\, \in \, Z^{\times 2},\, q\, \neq \, 0}
\sum_{\partial \xi^{1} \, =\, \delta^{0} [0]\, +\, \delta^{0} [q]}
\mid (\tanh \beta E)^{\xi^{1}}\mid
\end{equation}
is absolutely convergent. By using the inequality (\ref{3.15}) we majorize the sum
(\ref{2.27}) by the absolutely convergent series. It is sufficient now to consider the
convergence of any term of the sum (\ref{2.27}).

Let us prove that for any positive number $L$
\begin{eqnarray}
\label{3.18}
\lim_{G \rightarrow Z^{\times 2}} (\# (VG))^{- k}
\sum_{p_{1}\, \neq \, \cdots \, \neq \, p_{2k},\, p_{i}\, \in \, P(G)}
W_{G}(\sum_{i\, =\, 1}^{2k} \delta^{0} [{\bf p}_{i}]) =
2^{- k}\frac{(2k)!}{k!}
\lim_{G \rightarrow Z^{\times 2}} (\# (VG))^{- k} \nonumber \\
\sum_{{p_{1}\, \neq \, \cdots \, \neq \, p_{2k},\, p_{i}\, \in \, P(G);}
\atop {|p_{2i}\, -\, p_{2j}|\, >\, L,\, 1\, \leq \, i\, <\, j\, \leq \, k}}
\sum_{{\partial \xi_{i}^{1} \, =\, \delta^{0} [p_{2i - 1}]\, +\,
\delta^{0} [p_{2i}],}\atop {i\, =\, 1,...,k}}
\left( \prod_{i\, =\, 1}^{k} (\tanh \beta E)^{\xi_{i}^{1}}\right)
(Z_{r,G})^{- 1}Z_{r,G^{\prime}} .
\end{eqnarray}
In the relation (\ref{3.18}) we use the same notation as in the relation
(\ref{3.11}).

In order to prove the equality (\ref{3.18}) it is sufficient to prove
the following equality
\begin{eqnarray}
\label{3.19}
\lim_{G \rightarrow Z^{\times 2}} (\# (VG))^{- k}
\sum_{{p_{1}\, \neq \, \cdots \, \neq \, p_{2k},\, p_{i}\, \in \, P(G);}
\atop {|p_{2l}\, -\, p_{2j}|\, \leq \, L}}
\sum_{{\partial \xi_{i}^{1} \, =\, \delta^{0} [p_{2i - 1}]\, +\,
\delta^{0} [p_{2i}],}\atop {i = 1,...,k}} \nonumber \\
\left( \prod_{i\, =\, 1}^{k} \mid (\tanh \beta E)^{\xi_{i}^{1}}\mid \right)
(Z_{r,G})^{- 1}Z_{r,G^{\prime}} = 0
\end{eqnarray}
for any positive number $L$ and for any fixed integers $1 \leq l < j \leq k$.
In view of the estimate (\ref{3.4}) the equality (\ref{3.19}) follows
from the equality
\begin{equation}
\label{3.20}
\lim_{G \rightarrow Z^{\times 2}} (\# (VG))^{- k}
\sum_{{p_{1}\, \neq \, \cdots \, \neq \, p_{2k},\, p_{i}\, \in \, P(G);}
\atop {|p_{2l}\, -\, p_{2j}|\, \leq \, L}}
\sum_{{\partial \xi_{i}^{1} \, =\, \delta^{0} [p_{2i - 1}]\, +\,
\delta^{0} [p_{2i}],}\atop {i = 1,...,k}}
\prod_{i\, =\, 1}^{k} \mid (\tanh \beta E)^{\xi_{i}^{1}}\mid = 0.
\end{equation}
In order to prove the equality (\ref{3.20}) we majorize the sum (\ref{3.20})
by the similar sum where the second summing runs all non -- oriented
paths $\xi_{1}^{1}, ...,\xi_{k}^{1}$ on the lattice ${\bf Z}^{\times 2}$
connecting the vertices ${\bf p}_{2i - 1}$ and ${\bf p}_{2i}, i = 1,...,k$
and self -- intersecting transversally. Since the interaction energy
$E(\{ {\bf p},{\bf e}\} )$ depends only on the direction (horizontal or
vertical) of the edge $\{ {\bf p},{\bf e}\}$, then in view of the definition
(\ref{2.33})
\begin{eqnarray}
\label{3.21}
\sum_{{p_{1}\, \neq \, \cdots \, \neq \, p_{2k},\, p_{i}\, \in \, P(G);}
\atop {|p_{2l}\, -\, p_{2j}|\, \leq \, L}}
\sum_{{\partial \xi_{i}^{1} \, =\, \delta^{0} [p_{2i - 1}]\, +\,
\delta^{0} [p_{2i}],}\atop {i = 1,...,k}}
\prod_{i\, =\, 1}^{k} \mid (\tanh \beta E)^{\xi_{i}^{1}}\mid = \nonumber \\
\sum_{{q_{i}\, \in \, Z^{\times 2},\, q_{i}\, \neq \, 0,}
\atop {i\, =\, 1,...,k}}
\sum_{{\partial \xi_{i}^{1} \, =\, \delta^{0} [0]\, +\, \delta^{0} [q_{i}],}
\atop {i = 1,...,k}}
\prod_{i\, =\, 1}^{k} \mid (\tanh \beta E)^{\xi_{i}^{1}}\mid
\sum_{{p_{1}\, \neq \, \cdots \, \neq \, p_{2k},\,
p_{2i}\, -\, p_{2i - 1}\, =\, q_{i},\, p_{i}\, \in \, P(G);}
\atop {|p_{2l}\, -\, p_{2j}|\, \leq \, L}} 1.
\end{eqnarray}
It is easy to prove the inequality
\begin{equation}
\label{3.22}
\sum_{{p_{1}\, \neq \, \cdots \, \neq \, p_{2k},\,
p_{2i}\, -\, p_{2i - 1}\, =\, q_{i},\, p_{i}\, \in \, P(G);}
\atop {|p_{2l}\, -\, p_{2j}|\, \leq \, L}} 1
\leq (2L + 1)(\# (VG))^{k - 1}.
\end{equation}
The absolute convergence of the series (\ref{3.16}), the equality
(\ref{3.21}) and the estimate (\ref{3.22}) imply the equality (\ref{3.20}).
The equality (\ref{3.20}) implies the equalities (\ref{3.19}), (\ref{3.18}).

We denote by $|\xi^{1} |$ the total number of the edges
$\{ {\bf p},{\bf e}\} $ at which the non -- oriented path $\xi^{1}$
takes the value $1$. The absolute convergence of the series (\ref{3.16})
implies that for any positive number $\epsilon$ there is the sufficiently
large number $\frac{1}{4} L$ such that
\begin{equation}
\label{3.23} \sum_{q\, \in \, Z^{\times 2},\, q\, \neq \, 0,} \sum_{{\partial \xi^{1}
\, =\, \delta^{0} [0]\, +\, \delta^{0} [q],} \atop {|\xi^{1} |\, >\, \frac{1}{4} L}}
\mid (\tanh \beta E)^{\xi^{1}}\mid \, < \epsilon .
\end{equation}
The estimates (\ref{3.4}), (\ref{3.14}), (\ref{3.23}) imply the
following estimate
\begin{eqnarray}
\label{3.24}
(\# (VG))^{- k} \mid
\sum_{{p_{1}\, \neq \, \cdots \, \neq \, p_{2k},\, p_{i}\, \in \, P(G);}
\atop {|p_{2i}\, -\, p_{2j}|\, >\, L,\, 1\, \leq \, i\, <\, j\, \leq \, k}}
\sum_{{\partial \xi_{i}^{1} \, =\, \delta^{0} [p_{2i - 1}]\, +\,
\delta^{0} [p_{2i}],} \atop {i\, =\, 1,...,k}}
\left( \prod_{i\, =\, 1}^{k} (\tanh \beta E)^{\xi_{i}^{1} }\right)
(Z_{r,G})^{- 1}Z_{r,G^{\prime}} - \nonumber \\
\sum_{{p_{1}\, \neq \, \cdots \, \neq \, p_{2k},\, p_{i}\, \in \, P(G);} \atop
{|p_{2i}\, -\, p_{2j}|\, >\, L,\, 1\, \leq \, i\, <\, j\, \leq \, k}} \sum_{{\partial
\xi_{i}^{1}\, =\, \delta^{0} [p_{2i - 1}]\, +\, \delta^{0} [p_{2i}],} \atop
{|\xi_{i}^{1} |\, \leq \, \frac{1}{4} L,\, i\, =\, 1,...,k}} \left( \prod_{i\, =\,
1}^{k} (\tanh \beta E)^{\xi_{i}^{1} }\right) (Z_{r,G})^{- 1}Z_{r,G^{\prime}} \mid \, <
\epsilon C
\end{eqnarray}
where the constant $C$ is independent of the graph
$G(M^{\prime}_{1},M^{\prime}_{2};M_{1},M_{2})$.

Let us consider the limit
\begin{eqnarray}
\label{3.25}
\lim_{G \rightarrow Z^{\times 2}} (\# (VG))^{- k}
\sum_{{p_{1}\, \neq \, \cdots \, \neq \, p_{2k},\, p_{i}\, \in \, P(G);}
\atop {|p_{2i}\, -\, p_{2j}|\, >\, L,\, 1\, \leq \, i\, <\, j\, \leq \, k}}
\sum_{{\partial \xi_{i}^{1}\, =\, \delta^{0} [p_{2i - 1}]\, +\,
\delta^{0} [p_{2i}],}
\atop {|\xi_{i}^{1} |\, \leq \, \frac{1}{4} L,\, i\, =\, 1,...,k}}
\nonumber \\
\left( \prod_{i\, =\, 1}^{k} (\tanh \beta E)^{\xi_{i}^{1} }\right)
(Z_{r,G})^{- 1}Z_{r,G^{\prime}} .
\end{eqnarray}
The relation (\ref{2.55}) implies
\begin{eqnarray}
\label{3.26}
(Z_{r,G})^{- 1}Z_{r,G^{\prime}} =
\exp \{ \frac{1}{2} \sum_{i\, =\, 1}^{k} \sum_{{C\, \in \, CRC(G),}\atop
{||C||\, \cap \, ||j(\xi_{i}^{1})||\, \neq \, \emptyset }}
|C|^{- 1}\exp \{ \frac{i}{2} \phi (C)\} (\tanh \beta E)^{C}\}
\times \nonumber \\
\exp \{ - \frac{1}{2} \sum_{{C\, \in \, CRC(G),}\atop
{||C||\, \cap \, \cup_{i} ||j(\xi_{i}^{1})||\, \neq \, \emptyset}}
(n(C;\xi ) - 1)|C|^{- 1}\exp \{ \frac{i}{2} \phi (C)\} (\tanh \beta E)^{C}\}
\end{eqnarray}
where in the second multiplier the sum extends over the set of all
completely reduced closed paths having the common edges with at least two
non -- oriented paths $\xi_{i}^{1}, i = 1,...,k$. $n(C;\xi )$ is the number
of the non -- oriented paths $\xi_{1}^{1}, ...,\xi_{k}^{1}$ with which
the completely reduced closed path $C$ has the common edges. Since
$|{\bf p}_{2i} - {\bf p}_{2j}| > L$, $1 \leq i < j \leq k$;
$|\xi_{i}^{1}| \leq \frac{1}{4} L$, $i = 1,...,k$, the length of such
completely reduced closed path is more than $L$. In view of the
estimate (\ref{2.37}) the series
\begin{equation}
\label{3.27}
\sum_{{C\, \in \, CRC(G),}\atop
{||C||\, \cap \, \cup_{i} ||j(\xi_{i}^{1})||\, \neq \, \emptyset}}
(n(C;\xi ) - 1)|C|^{- 1}\exp \{ \frac{i}{2} \phi (C)\} (\tanh \beta E)^{C}
\end{equation}
tends to zero when $L \rightarrow \infty $. By choosing the sufficiently
large number $L$ we can approximate the limit (\ref{3.25}) by the
following limit
\begin{eqnarray}
\label{3.28}
\lim_{G \rightarrow Z^{\times 2}} (\# (VG))^{- k}
\sum_{{p_{1}\, \neq \, \cdots \, \neq \, p_{2k},\, p_{i}\, \in \, P(G);}
\atop {|p_{2i}\, -\, p_{2j}|\, >\, L,\, 1\, \leq \, i\, <\, j\, \leq \, k}}
\sum_{{\partial \xi_{i}^{1}\, =\, \delta^{0} [p_{2i - 1}]\, +\,
\delta^{0} [p_{2i}],}
\atop {|\xi_{i}^{1} |\, \leq \, \frac{1}{4} L,\, i\, =\, 1,...,k}}
\nonumber \\
\prod_{i\, =\, 1}^{k} \left( (\tanh \beta E)^{\xi_{i}^{1} }
\exp \{ \frac{1}{2} \sum_{{C\, \in \, CRC(G),}\atop
{||C||\, \cap \, ||j(\xi_{i}^{1})||\, \neq \, \emptyset }}
|C|^{- 1}\exp \{ \frac{i}{2} \phi (C)\} (\tanh \beta E)^{C}\} \right) .
\end{eqnarray}
Due to the equality (\ref{2.55})
\begin{equation}
\label{3.29}
\exp \{ \frac{1}{2} \sum_{{C\, \in \, CRC(G),}\atop
{||C||\, \cap \, ||j(\xi_{i}^{1})||\, \neq \, \emptyset }}
|C|^{- 1}\exp \{ \frac{i}{2} \phi (C)\} (\tanh \beta E)^{C}\} =
(Z_{r,G})^{- 1}Z_{r,G \setminus ||j(\xi_{i}^{1})||}
\end{equation}
where the graph $G \setminus ||j(\xi_{i}^{1})||$ is obtained from the graph
$G(M^{\prime}_{1},M^{\prime}_{2};M_{1},M_{2})$ by deleting all edges from
the support $||j(\xi_{i}^{1})||$. We substitute the relation (\ref{3.29})
into the expression (\ref{3.28}). By making use of the estimate (\ref{3.4})
and the equality (\ref{3.20}) it is possible to prove that the expression
(\ref{3.28}) is equal to
\begin{eqnarray}
\label{3.30}
\lim_{G \rightarrow Z^{\times 2}} (\# (VG))^{- k}
\sum_{p_{1}\, \neq \, \cdots \, \neq \, p_{2k},\, p_{i}\, \in \, P(G)}
\sum_{{\partial \xi_{i}^{1}\, =\, \delta^{0} [p_{2i - 1}]\, +\,
\delta^{0} [p_{2i}],}
\atop {|\xi_{i}^{1} |\, \leq \, \frac{1}{4} L,\, i\, =\, 1,...,k}}
\nonumber \\
\prod_{i\, =\, 1}^{k} \left( (\tanh \beta E)^{\xi_{i}^{1} }
\exp \{ \frac{1}{2} \sum_{{C\, \in \, CRC(G),}\atop
{||C||\, \cap \, ||j(\xi_{i}^{1})||\, \neq \, \emptyset }}
|C|^{- 1}\exp \{ \frac{i}{2} \phi (C)\} (\tanh \beta E)^{C}\} \right) =
\nonumber \\
\lim_{G \rightarrow Z^{\times 2}} ( (\# (VG))^{- 1}
\sum_{p_{1}\, \neq \, p_{2},\, p_{i}\, \in \, P(G)}
\sum_{{\partial \xi^{1}\, =\, \delta^{0} [p_{1}]\, +\, \delta^{0} [p_{2}],}
\atop {|\xi^{1} |\, \leq \, \frac{1}{4} L}}
\nonumber \\
(\tanh \beta E)^{\xi^{1}}\exp \{ \frac{1}{2} \sum_{{C\, \in \, CRC(G),}\atop
{||C||\, \cap \, ||j(\xi^{1})||\, \neq \, \emptyset }}
|C|^{- 1}\exp \{ \frac{i}{2} \phi (C)\} (\tanh \beta E)^{C}\} )^{k}
\end{eqnarray}
where $\xi^{1}$ runs all non -- oriented paths on the lattice
$G(M^{\prime}_{1},M^{\prime}_{2};M_{1},M_{2})$ connecting the vertices
${\bf p}_{1}$ and ${\bf p}_{2}$. These non -- oriented paths self --
intersect transversally. The length $|\xi^{1}|$ of the non -- oriented
path $\xi^{1}$ is less than $\frac{1}{4} L$. When $L \rightarrow \infty$
the equalities (\ref{3.18}), (\ref{3.30}) and the estimate (\ref{3.24}) imply
\begin{eqnarray}
\label{3.31}
\lim_{G \rightarrow Z^{\times 2}} (\# (VG))^{- k}
\sum_{p_{1}\, \neq \, \cdots \, \neq \, p_{2k},\, p_{i}\, \in \, P(G)}
W_{G}(\sum_{i\, =\, 1}^{2k} \delta^{0} [{\bf p}_{i}]) =
\nonumber \\
2^{- k}\frac{(2k)!}{k!}
\lim_{G \rightarrow Z^{\times 2}} ((\# (VG))^{- 1}
\sum_{p_{1}\, \neq \, p_{2},\, p_{i}\, \in \, P(G);}
W_{G}(\sum_{i\, =\, 1}^{2} \delta^{0} [{\bf p}_{i}]))^{k}.
\end{eqnarray}
Here we used the expression (\ref{2.56}) for the correlation function
$W_{G}(\sum_{i\, =\, 1}^{2} \delta^{0} [{\bf p}_{i}])$.

Since the interaction energy $E(\{ {\bf p},{\bf e}\} )$ depends only on
the direction (horizontal or vertical) of the edge $\{ {\bf p},{\bf e}\}$,
the definitions (\ref{2.33}), (\ref{2.43}) and the equality (\ref{2.56})
imply
\begin{eqnarray}
\label{3.32}
\lim_{G \rightarrow Z^{\times 2}} (\# (VG))^{- 1}
\sum_{p_{1}\, \neq \, p_{2},\, p_{i}\, \in \, P(G)}
W_{G}(\sum_{i\, =\, 1}^{2} \delta^{0} [{\bf p}_{i}]) =
\sum_{p\, \in \, Z^{\times 2},\, p\, \neq \, 0}
\sum_{\partial \xi^{1} \, =\, \delta^{0} [0]\, +\, \delta^{0} [p]}
\nonumber \\
(\tanh \beta E)^{\xi^{1}}\exp \{ \frac{1}{2}
\sum_{{C\, \in \, CRC(Z^{\times 2}),}\atop
{||C||\, \cap \, ||j(\xi^{1})||\, \neq \, \emptyset }}
|C|^{- 1}\exp \{ \frac{i}{2} \phi (C)\} (\tanh \beta E)^{C}\}
\times \nonumber \\
\lim_{G \rightarrow Z^{\times 2}} (\# (VG))^{- 1}N_{G}(\xi^{1} )
\end{eqnarray}
where the number $N_{G}(\xi^{1} )$ is the total number of shifted
non -- oriented paths $\xi^{1}$ on the graph
$G(M^{\prime}_{1},M^{\prime}_{2};M_{1},M_{2})$. If
$|\xi^{1}| \leq M_{i} - M_{i}^{\prime}$, $i = 1,2$, the following
estimates are valid
\begin{equation}
\label{3.33}
\prod_{i\, =\, 1}^{2} (M_{i} - M_{i}^{\prime} - |\xi^{1}|) \leq
N_{G}(\xi^{1} ) \leq \prod_{i\, =\, 1}^{2} (M_{i} - M_{i}^{\prime}).
\end{equation}
We note that
\begin{equation}
\label{3.34}
\# (VG) = \prod_{i\, =\, 1}^{2} (M_{i} - M_{i}^{\prime} + 1).
\end{equation}
The equality (\ref{3.29}) and the estimates (\ref{2.37}), (\ref{3.4})
imply the absolute convergence of the series
\begin{eqnarray}
\label{3.35} \sum_{p\, \in \, Z^{\times 2},\, p\, \neq \, 0} W_{Z^{\times
2}}(\delta^{0} [{\bf 0}] + \delta^{0} [{\bf p}]) = \sum_{p\, \in \, Z^{\times 2},\, p\,
\neq \, 0} \sum_{\partial \xi^{1} \, =\, \delta^{0} [0]\, +\, \delta^{0} [p]}
\nonumber \\
(\tanh \beta E)^{\xi^{1}}\exp \{ \frac{1}{2}
\sum_{{C\, \in \, CRC(Z^{\times 2}),}\atop
{||C||\, \cap \, ||j(\xi^{1})||\, \neq \, \emptyset }}
|C|^{- 1}\exp \{ \frac{i}{2} \phi (C)\} (\tanh \beta E)^{C}\}.
\end{eqnarray}
Thus it follows from the equalities (\ref{3.32}), (\ref{3.34}), (\ref{3.35})
and the estimates (\ref{3.33}) that
\begin{equation}
\label{3.36} \lim_{G \rightarrow Z^{\times 2}} (\# (VG))^{- 1} \sum_{p_{1}\, \neq \,
p_{2},\, p_{i}\, \in \, P(G)} W_{G}(\sum_{i\, =\, 1}^{2} \delta^{0} [{\bf p}_{i}]) =
\sum_{p\, \in \, Z^{\times 2},\, p\, \neq \, 0} W_{Z^{\times 2}}(\delta^{0} [{\bf 0}] +
\delta^{0} [{\bf p}]).
\end{equation}
The inequality (\ref{3.15}) and the equalities (\ref{3.31}), (\ref{3.36}) imply the
equality (\ref{3.9}). The proposition is proved.

It follows from the equalities (\ref{2.23}), (\ref{3.9}) and the inequality
(\ref{3.15}) that
\begin{equation}
\label{3.37} \lim_{G \rightarrow \infty } m_{G}(H) = \beta H(1 + \sum_{p\, \in \,
Z^{\times 2},\, p \, \neq \, 0} W_{Z^{\times 2}}(\delta^{0} [{\bf 0}] + \delta^{0}
[{\bf p}])).
\end{equation}

Let us rewrite the formula (\ref{1.8}). Due to (\cite{5}, Chapter III, formula (3.1))
we define
\begin{equation}
\label{3.38} <\sigma_{m} \sigma_{n} >_{N} = (Z_{N}(E,H))^{- 1}\sum_{{\sigma_{k} \, =\,
\pm 1, }\atop{k\, =\, 1,...,N}} \sigma_{m} \sigma_{n} \exp \{ - \beta \bar{H} (\sigma)\}
\end{equation}
where the integers $1 \leq m \leq N$, $1 \leq n \leq N$ and the sums $\bar{H} (\sigma)$,
$Z_{N}(E,H)$ are given by the relations (\ref{1.1}), (\ref{1.2}). By keeping $m - n$
fixed we have due to (\cite{5}, Chapter III, formula (3.5))
\begin{equation}
\label{3.39} \lim_{N \rightarrow \infty} <\sigma_{m} \sigma_{n} >_{N}\mid_{H = 0} =
(\tanh \beta E)^{|m - n|}.
\end{equation}
Let the integer $N$ be even. For any integers $m$ and $n$ we define the correlation
function
\begin{equation}
\label{3.40} W_{Z}(\delta^{0} [m] + \delta^{0} [n]) = \lim_{N \rightarrow \infty}
<\sigma_{\frac{N}{2} + m} \sigma_{\frac{N}{2} + n}>_{N}\mid_{H = 0}.
\end{equation}
The relations (\ref{3.39}), (\ref{3.40}) imply
\begin{equation}
\label{3.41} \sum_{n\, \in \, Z,\, n \, \neq \, 0} W_{Z}(\delta^{0} [0] + \delta^{0}
[n]) = \frac{2\tanh \beta E}{1 - \tanh \beta E} = e^{2\beta E} - 1.
\end{equation}
The substitution of the relation (\ref{3.41}) into the equality (\ref{1.8}) gives
\begin{equation}
\label{3.42} m(E,H) = \beta H(1 + \sum_{n\, \in \, Z,\, n \, \neq \, 0}
W_{Z}(\delta^{0} [0] + \delta^{0} [n])).
\end{equation}
The equalities (\ref{3.37}) and (\ref{3.42}) have the similar form.

\end{document}